\documentclass[psamfonts, 12pt]{article}
\usepackage[mathscr]{eucal}
\usepackage{amsthm,amsmath,amssymb,latexsym,amscd}
\newtheorem{lem}{Lemma}[section]
\newtheorem{dfn}[lem]{Definition}
\newtheorem{prop}[lem]{Proposition}
\newtheorem{teo}[lem]{Theorem}

\newtheorem{cor}[lem]{Corollary}
\newtheorem{exam}[lem]{Example}
\font\cj=msbm10
at 12pt \font\eightmsb=msbm10 at 8pt \font\sixmsb=msbm10 at 6pt
\title{Non-Conservative Minimal Quantum Dynamical Semigroups}
\author{R. Quezada Batalla\thanks{ Partially supported by {\sc conacyt}.}}
\date{}
\newfam\msbfam
\textfont\msbfam=\cj \scriptfont\msbfam=\eightmsb
\scriptscriptfont\msbfam=\sixmsb

\font\euf=eufm10 at 12pt

\def\h{\mbox{\euf h}}
\def\ka{\mbox{\euf k}}

\def\dom{\mbox{\rm dom}}
\def\tr{\mbox{\rm tr}}
\def\span{\mbox{\rm span}}

\def\k{\mathbb C}

\def\tr{\mbox{\rm tr}}
\mathchardef\ssneq="3B28
\mathchardef\pez="3B6E

\let\dps=\displaystyle
\def\m@th{\mathsurround=\z@}

\begin{document}

\maketitle

\begin{abstract}
Necessary and sufficient conditions for non-conservativity of a
class of quantum dynamical semigroups are given. Extensions of
well known criteria for conservativity are obtained and
interesting connections of the conservativity problem with the von
Neuman´s Theory of the defect indices for symmetric operators are
studied.
\end{abstract}

\section{Introduction}
The concept of quantum dynamical semigroup (qds) has become a
fundamental notion in the theory of quantum Markov processes. The
theory of qds has been intensively studied in recent years, laying
special emphasis to the so called minimal quantum dynamical
semigroup as well as to sufficient conditions to ensure its
conservativity (markovianity or unitality) [2], [5]. This approach
has yields to distinguish a class of minimal conservative qds.
Much less attention has received the class of non-conservative
qds, nevertheless the study of this class is important both from
the mathematical point of view as well as for applications in
models of quantum physics.

The main aim of this work is to describe a class of
non-conservative minimal qds that naturally arises when a
necessary condition for conservativity is not satisfied.In the
case when the CP part of the formal generator is zero, a more
careful study of the formal generators of this class of qds
permits one to observe interesting connections with the von
Neumann´s theory of the defect indices of symmetric operators.

In section 2 we give the necessary definitions. Section 3 contains
several criteria for non-conservativity or explosion of the class
of qds introduced in section 2, in particular, Theorems 3.1 and
3.3 entend some criteria of A.M. Chebotarev for conservativity and
Corollary 3.5 extends a well known criterion of E.B. Davies. In
Section 4 we provide several examples and show the connections of
the conservativity problem for the class of minimal qds introduced
in Section 2 with the von Neumann Theory of the defect indices of
symmetric operator, in the case when the CP part of the formal
generator is zero.

\section{Preliminaries}
Along this work $\h$ will denote a separable complex Hilbert space
with the inner product $\langle \;,\;\rangle $ and the norm
$\|\cdot\|$, ${\cal B}={\cal B}(\h)$ will denote the von Neumann
algebra of all bounded linear operators in $\h$ and
$\|\cdot\|_\infty$ will denote the norm in this space.

\begin{dfn}
 A quantum dynamical semigroup on ${\cal B}$ is a semigroup
$P=(P_t)_{t\ge 0}$ of bounded operators in ${\cal B}$ with the
following properties
\begin{itemize}
\item[(a)] Complete Positivity (CP). $P_t$ is completely positive for every $t\ge 0$,
i.e. for every pair of finite sequences $(x_i), (y_j)$ in ${\cal
B}$ $$\sum_{i,j}y^*_iP_t(x^*_ix_j)y_j\ge 0.$$
\item[(b)] (Normality or $\sigma$-weak continuity). For every increasing
net $(x_\alpha)$ of positive elements in ${\cal B}$ with an upper
bound we have  $$P_t(\sup_\alpha x_\alpha)=\sup_\alpha
P_t(x_\alpha)$$ for every $t\ge 0$.
\item[(c)] (Ultraweak or $*$-weak continuity in $t$). For every
trace class operator and every $x\in{\cal B}$ we have that
$$\lim_{t\to 0^+ }tr (\rho P_t(x))=tr(\rho x).$$
\item[(d)] $P_t(I)\le I$ for all $t\ge 0$.
\end{itemize}
\end{dfn}
A qds $(P_t)_{t\ge 0}$ is conservative (markovian or unital) if
$P_t(I)=I$, for all $t\ge 0$.

If a conservative qds is uniformly continuous $\dps
\left(\lim_{t\to 0^+}\sup_{{\|x\|_\infty}=1}\|P_t(x)-x\|_\infty
=0\right)$, then its infinitesimal generator is a bounded linear
operator ${\cal L}:{\cal B}\to{\cal B}$ and there exist a CP
normal bounded map $\phi:{\cal B}\to{\cal B}$ and a bounded self
adjoint operator $H$ such that $${\cal
L}(x)=\phi(x)-G^*x-xG\eqno(2.1)$$ with $G=(1/2)\phi(I)-iH$. And
conversely any linear operator ${\cal L}$ with the structure (2.1)
is the infinitesimal generator of a uniformly continuous
conservative qds. This is an important result due to Linblad and
Gorini-Kossakowski-Sudarshan, see [5] and the references therein.

In this work we shall consider unbounded formal generators ${\cal
L}$ that associates with every $x\in{\cal B}$ an unbounded
sesquilinear form with the structure $${\cal
L}(x)[u,v]=\phi(x)[u,v]-\langle Gu,xv\rangle -\langle u,xGv\rangle
,$$ $u,v\in\dom G$, where
\begin{itemize}
\item[(i)] $-G$ is the generator of a $C_o$-semigroup of
contractions in $\h$, $(W_t)_{t\ge 0}$.
\item[(ii)] $\phi$ is a linear CP and normal map, i.e., for every
$x\in{\cal B}$, $\phi(x)$ is a sesquilinear form defined on $\dom
G\times\dom G$ such that
\begin{itemize}
\item[(ii.1)] (CP): for any pair of finite sequences $(u_i)\in\dom G$
and $(x_i)\subset{\cal B}$ we have that
$$\sum_{i,j}\phi(x^*_ix_j)[u_i, u_j]\ge 0.$$
\item[(ii.2)] For every $u\in\dom G$, $\phi(\cdot)[u]$ is a normal
linear functional on ${\cal B}$, i.e., for any increasing net
$(x_\alpha)$ of positive elements of ${\cal B}$ with an upper
bound, $$\phi\left(\sup_\alpha
x_\alpha\right)[u]=\sup_\alpha\phi(x_\alpha)[u],$$ where
$\phi(\cdot)[u]=\phi(\cdot)[u,u]$ is the quadratic form associated
with $\phi(\cdot)$.
\end{itemize}
\item[(iii)] the estimate $$0\le\phi(I)[u]\le Re\langle Gu, u\rangle $$ or
equivalently $${\cal L}(I)[u]\le 0,$$ holds for every $u\in\dom
G$.
\end{itemize}
Conditions (i)-(iii) are sufficient to construct a minimal qds
$\left(P^{\min}_t\right)_{t\ge 0}$ that satisfies the master
equation $$\frac{d}{dt}\left\langle u,P^{\min}_t(x)v\right\rangle
={\cal L}\left(P^{\min}_t(x)\right)[u,v], \quad
P^{\min}_0(x)=x,\eqno(2.2)$$ $u,v\in\dom G,\;\;x\in{\cal B},$
which it is shown to be equivalent with the integral equation
$$\frac{d}{dt}\left\langle u,P^{\min}_t(x)v\right\rangle =\langle
u,W^*_txW_tv\rangle
+\int^t_0d\tau\phi\left(P^{\min}_\tau(x)\right)[W_{t-\tau}u,W_{t-\tau}v],$$
$u,v\in\dom G, \; x\in{\cal B}$.

The minimal qds $(P^{\min}_t)_{t\ge 0}$ is not necessarily
conservative and the problem of finding necessary and sufficient
conditions for its conservativity has received the attention of
the people working in this topic. A. M. Chebotarev [2] and F.
Fagnola [5] have found necessary and  sufficient or only
sufficient conditions for the conservativity of the class of
minimal qds whose formal generator satisfy the additional
necessary condition
\begin{itemize}
\item[(iii')] $\dps{\cal L}(I)[u,v]=0, \quad \forall u,v\in\dom
G,$
\end{itemize}
which is an stronger form of (iii).

Our aim in this work is to study necessary and sufficient
conditions for non-conservativity of the class of minimal qds
whose formal generator satisfy only the conditions (i)-(iii). As a
corollary we will obtain well known criteria of A.M. Chebotarev
and E.B. Davies for conservativity.

\section{Criteria for explosion of the minimal qds.}
Our analysis is based on the quantity (or observable) ${\cal
E}_t(I)$, which we call ``probability for explosion at time $t$".
This quantity is defined as the positive bounded operator given by
$${\cal E}_t(I):=I-P^{\min}_t(I).$$ For the construction of
$(P^{\min}_t)_{t\ge 0}$ it is used the following iterative scheme:
$$P^{(1)}_t(x)[u,v]=\langle u,W^*_txW_tv\rangle $$ and
$$P^{(n)}_t(x)[u,v]=\langle u,W^*_txW_tv\rangle
+\int^t_0d\tau\phi\left(P^{(n-1)}_\tau(x)\right)[W_{t-\tau}u,W_{t-\tau}v],$$
for $u$, $\;v\in\dom G$, $x\in{\cal B}$ and $t\ge 0$ fixed.

It is proved in [2] and [5], that for $x\ge 0$ the sequence of
positive operators $\left(P^{(n)}_t(x)\right)_{n\ge 0}$, $t\ge 0$
fixed, is increasing and bounded, therefore there exists
$$P^{\min}_t(x)=\sup_nP^{(n)}_t(x).$$ which is a solution of the
master equation (2.2).

 Let us consider the sequence $\left({\cal
E}^{(n)}_t(I)\right)_{n\ge 0}$, $t\ge 0$ fixed, defined as $${\cal
E}^{(n)}_t(I):=I-P^{(n)}_t(I).$$ We have that $${\cal
E}^{(1)}_t(I)[u,v]=\langle u,(I-W^*_tW_t)v\rangle , \;\;
u,v\in\dom G,$$ and for $n\ge 2$ we have $${\cal
E}^{(n)}_t(I)[u]=\langle u,{\cal
E}^{(1)}_t(I)u\rangle+\int^t_0d\tau\phi\big({\cal
E}^{(n-1)}_\tau(I)\big)[W_{t-\tau}u]-\int^t_0d\tau\phi(I)[W_{t-\tau}u],$$
for $u\in\dom G$.

By performing a Laplace transform we obtain
\begin{eqnarray*}
\tilde{\cal E}^{(n)}_\lambda(I)[u]&=&\tilde{\cal
E}^{(1)}_\lambda(I)[u]+\int^\infty_0dte^{-\lambda
t}\phi\big(\tilde{\cal
E}^{(n-1)}_\lambda(I)\big)[W_tu]-\frac{1}{\lambda}Q_\lambda(I)[u]\\
&=&\left(\tilde{\cal E}^{(1)}_\lambda(I)-\frac{1}{\lambda}
Q_\lambda(I)\right)[u]+Q_\lambda\left(\tilde{\cal
E}^{(n-1)}_\lambda(I)\right)[u],\end{eqnarray*} $u\in\dom G$,
where $\tilde{\cal E}^{(n)}_\lambda(I)$ is the Laplace transform
of the sesquilinear form associated with ${\cal E}^{(n)}_t(I)$ and
$$Q_\lambda(x)[u]=\int^\infty_0dte^{-\lambda t}\phi(x)[W_tu].$$ It
can be shown that $\tilde{\cal E}^{(n)}_\lambda(I)$ and
$Q_\lambda(x)$ are bounded sesquilinear forms in $\h$ and we shall
denote by the same symbols the corresponding bounded operators.

 Therefore $$\tilde{\cal
E}^{(n)}_\lambda(I)[u]=\ell_\lambda(I)[u]+Q_\lambda
\left(\tilde{\cal E}^{(n-1)}_\lambda(I)\right)[u],\quad u\in\dom
G,$$ with $$\begin{array}{lll}&&\dps
\ell_\lambda(I)[u]=\big(\tilde{\cal
E}^{(1)}_\lambda(I)-\frac{1}{\lambda}
Q_\lambda(I)\big)[u]=\int^\infty_0dte^{-\lambda t}\left[\langle
u,(I-W^*_tW_t)u\rangle \right.\\ &&\dps
\left.-\int^t_0d\tau\phi(I)[W_{t-\tau}u]\right]=\int^\infty_0dte^{-\lambda
t}\int^t_0d\tau\left(\frac{d}{d\tau}\|W_{t-\tau}u\|^2
-\phi(I)[W_{t-\tau}u]\right)\\ &=&\dps \int^\infty_0dte^{-\lambda
t}\int^t_0d\tau\bigg(\langle GW_{t-\tau}u,W_{t-\tau}u\rangle
+\langle
W_{t-\tau}u,GW_{t-\tau}u\rangle-\phi(I)[W_{t-\tau}u]\bigg)
\\ &=&\dps \int^\infty_0dt e^{-\lambda t}\left(-\int^t_0d\tau{\cal
L}(I)[W_{t-\tau}u]\right).
\end{array}$$
Notice that $\ell_\lambda(I)$ is a positive bounded sesquilinear
form since ${\cal L}(I)[W_{t-\tau}u]\le 0$, $u\in\dom \;G$.

Consequently we obtain: \begin{eqnarray*} \tilde{\cal
E}^{(n)}_\lambda(I)[u]&=&\ell_\lambda(I)[u]+Q_\lambda\left(\ell_\lambda(I)+Q_\lambda\left(\tilde{\cal
E}^{(n-2)}_\lambda(I)\right)\right)[u]\\
&=&\ell_\lambda(I)[u]+Q_\lambda\left(\ell_\lambda(I)\right)[u]+Q^2_\lambda\left(\ell_\lambda(I)+Q_\lambda\left(\tilde{\cal
E}^{(n-3)}_\lambda(I)\right)\right)[u]\\
&=&\ell_\lambda(I)[u]+Q_\lambda\big(\ell_\lambda(I)\big)[u]+Q^2_\lambda\big(\ell_\lambda(I)\big)[u]+\cdots
+Q^{n-2}_\lambda\big(\ell_\lambda(I)\big)\\&& +
Q^{n-1}_\lambda\left(\tilde{\cal
E}^{(1)}_\lambda(I)\right)=\sum^{n-2}_{k=0}Q^k_\lambda\left(\ell_\lambda(I)\right)[u]+Q^{n-1}_\lambda\left(\ell_\lambda(I)+\frac{1}{\lambda}Q_\lambda(I)\right)\\
\end{eqnarray*}
$$=\sum^{n-1}_{k=0}Q_\lambda\left(\ell_\lambda(I)\right)[u]+\frac{1}{\lambda}Q^n_\lambda(I),\eqno(3.1)$$
since $\dps\tilde{\cal
E}^{(1)}_\lambda(I)=\ell_\lambda(I)+\frac{1}{\lambda}Q_\lambda(I)$.

It is shown in [2], [5] that the sequence of positive operators
$\big(Q^n_\lambda(I)\big)_{n\ge 1}$ is convergent in $*$-weak and
strong sense. The sequence ${\cal E}^{(n)}_t(I)=I-P^{(n)}_t(I)\ge
0$ is a decreasing sequence of positive elements in ${\cal B}$,
hence the limit $\dps\lim_n{\cal E}^{(n)}_t(I)$ exists in $*$-weak
and strong sense. Therefore using the Lebesgue theorem on
dominated convergence we obtain $$\tilde{\cal
E}_\lambda(I)[u]=\int^\infty_0dte^{-\lambda t}\lim_n\langle
u,{\cal E}^{(n)}_t(I)u\rangle =\lim_n\langle u,\tilde{\cal
E}_\lambda(I)u\rangle ,$$ i.e, $\dps\tilde{\cal
E}_\lambda(I)=\lim_n\tilde{\cal E}^{(n)}_\lambda(I)$ in $*$-weak
and strong sense.

From (3.1) we obtain the following explicit formula for
$\tilde{\cal E}_\lambda(I)$: $$\tilde{\cal
E}_\lambda(I)=\frac{1}{\lambda}\lim_nQ^n_\lambda(I)+\sum_{n\ge
0}Q^n_\lambda\big(\ell_\lambda(I)\big),\eqno(3.2)$$ the limits
taken in $*$-weak or strong sense.

By $R^{\min}_\lambda$ we denote the resolvent map associated with
the qds $(P^{\min}_t)_{t\ge 0}$, i.e, for every $x\in{\cal B}$,
$R^{\min}_\lambda(x)$ is the operator in ${\cal B}$ defined by
means of the sesquilinear form
$$R^{\min}_\lambda(x)[u,v]=\int^\infty_0dte^{-\lambda t}\langle
u,P^{\min}_t(x)v\rangle, \quad u,v\in\h.$$

Therefore one has the following criterion for explosion or
non-conservativity of a minimal qds.

\begin{teo}
If ${\cal L}$ is a formal generator satisfying (i)-(iii), then the
following are equivalent
\begin{itemize}
\item[(i)] $\big(P^{\min}_t\big)_{t\ge 0}$ is non-conservative (or
explosive)
\item[(ii)] $\dps \ell_\lambda(I)\ne 0$ or
$\dps\lim_nQ^n_\lambda(I)\ne 0$.
\item[(iii)] $\dps R^{\min}_\lambda(I)<\frac{1}{\lambda}I$
\end{itemize}
\end{teo}

\noindent{\bf Proof.} The equivalence of (i) and (ii) follows
directly from (3.2).

Notice that for $\lambda >0$
\begin{eqnarray*}
R^{\min}_\lambda(I)[u]=\int^\infty_0dte^{-\lambda t}\langle
u,P^{\min}_t(I)u\rangle =\int^\infty_0dte^{-\lambda t}\|u\|^2-\\
-\int^\infty_0dte^{-\lambda t}\langle u,{\cal
E}_t(I)u\rangle=\frac{1}{\lambda}\|u\|^2-\tilde{\cal
E}_\lambda(I)[u],\end{eqnarray*} therefore
$R^{\min}_\lambda(I)<\frac{1}{\lambda}I$ if and only if
$\tilde{\cal E}_\lambda(I)> 0$. This proves that (i) and (iii) are
equivalent.\qed

\bigskip

 As a simple corollary we obtain Chebotarev's criterion for conservativity.

\begin{cor} If in addition ${\cal L}$ satisfies the condition
(iii)' then $\ell_\lambda(I)=0$ and $\dps\tilde{\cal
E}_\lambda(I)=\frac{1}{\lambda}\lim_nQ^n_\lambda(I)$. Hence
$(P^{\min}_t)_{t\ge 0}$ is conservative if and only if
$\dps\lim_nQ^n_\lambda(I)=0$.
\end{cor}

\bigskip

\noindent{\bf Proof.} Condition (iii)' implies that ${\cal
L}(I)[W_tu]=0$, $\forall t\ge 0$ and $u\in\dom\; G$. Therefore
$$\ell_\lambda(I)[u]=\int^\infty_0dte^{-\lambda
t}\left(-\int^t_0d\tau{\cal L}(I)[W_{t-\tau}u]\right)=0,\quad
\forall u\in\dom\; G.$$ This implies that $\ell_\lambda(I)=0$ as
an element of ${\cal B}$ since $\dom G$ is dense in $\h$. Hence,
it follows from (3.2) that $$\tilde{\cal
E}_\lambda(I)=\frac{1}{\lambda}\lim_nQ^n_\lambda(I)$$ in weak,
$*$-weak and strong sense.\qed

\bigskip

If ${\cal E}_t(x):=x-P^{\min}_t(x)$, $x\!\in\!{\cal B}$, then we
have from the master equation (2.2) that \begin{eqnarray*} {\cal
E}_t(x)[u,v]&=&-\int^t_0d\tau{\cal
L}\left(P^{\min}_\tau(x)\right)[u,v] =\\ &=& \int^t_0d\tau{\cal
L}\big({\cal E}_\tau(x)\big)[u,v]-t{\cal L}(x)[u,v],\quad
u,v\in\dom\;G.
\end{eqnarray*}

Hence with $x=I$ we obtain that $${\cal E}_t(I)[u,v]=-t{\cal
L}(I)[u,v]+\int^t_0d\tau{\cal L}\big({\cal E}_\tau(I)\big)[u,v],$$
$u,v\in\dom\; G$.

Assuming that

\medskip

\noindent (iv) ${\cal L}(I)\; [u,v]=0$, $\forall \; u,v\in
D\subset\dom G$, we obtain $D$ a dense subspace of $h$,  after
performing a Laplace transform, that $$\tilde{\cal
E}_\lambda(I)[u,v]=\frac{1}{\lambda}\int^\infty_0dte^{-\lambda
t}{\cal L}\left({\cal E}_t(I)\right)[u,v]=\frac{1}{\lambda}{\cal
L}\left(\tilde{\cal E}_\lambda(I)\right)[u,v],$$ $u,v\in D$.

\bigskip

Therefore $\tilde{\cal E}_\lambda(I)$ is a positive solution of
the equation. $${\cal L}(x)[u,v]=\lambda\langle u,xv\rangle,\;
u,v\in D,\; \lambda > 0,\; x\in{\cal B}.\eqno(3.3)$$ Notice that
(iv) is a weaker form of Chevotarev's condition (iii)'. The dense
subspace $D$ is not necessarily a core for $G$.
\begin{teo} Assume that ${\cal L}$ is a formal generator satisfying (i)-(iii) and (iv).
Then the following are equivalent
\begin{itemize}
\item[(i)] $\big(P^{\min}_t\big)_{t\ge 0}$ is non-conservative,
\item[(ii)] There exists a positive, bounded solution $x$ of (3.3) for some
\quad $\lambda >0.$
\end{itemize}
\end{teo}

\noindent{\bf Proof.} If $\left(P^{\min}_t\right)_{t\ge 0}$ is
non-conservative, equation (3.3) has the nontrivial solution $\dps
0 < x=\frac{\tilde{\cal E}_\lambda(I)}{\|\tilde{\cal
E}_\lambda(I)\|_\infty}$ for any $\lambda >0$.

\bigskip

Conversely, if $0 < x\le I$ is a positive bounded solution of
(3.3) for some $\lambda >0$, then  ${\cal L}(x)$ has a bounded
extension to the whole $\h$ and ${\cal L}(x)[u,v]=\lambda\langle
u, xv\rangle$ holds for every $u,v\in\dom G$, therefore for any
$u\in\dom\;G$  $$\begin{array}{lll}\dps e^{-\lambda
t}\phi(x)[W_tu]&=&\dps e^{-\lambda t}\big(\langle GW_tu,
xW_tu\rangle +\langle W_tu,xGW_tu\rangle + \lambda\langle
W_tu,xW_tu\rangle\big) \\ &=&\dps -\frac{d}{dt}e^{-\lambda
t}\langle W_tu,xW_tu\rangle .\end{array} \eqno(3.4)$$ hence
$$Q_\lambda(x)[u]=\int^\infty_0dte^{-\lambda
t}\phi(x)[W_tu]=-\int^\infty_0dt\frac{d}{dt}e^{-\lambda t}\langle
W_tu,xW_tu\rangle =\langle u,xu\rangle,$$ $u\in\dom\; G$, i.e.,
$$Q_\lambda(x)=x.\eqno(3.5)$$

Since $\phi(x)$ is positive and $$\phi(x)[W_tu]\le
\|x\|_\infty\phi(I)[W_tu]\le-\frac{d}{dt}\|W_tu\|^2,$$ we obtain
from (3.4) that $$-\frac{d}{dt}e^{-\lambda t}\langle
W_tu,xW_tu\rangle\le -e^{-\lambda t}\frac{d}{dt}\|W_tu\|^2, \quad
u\in\dom\; G.$$ Therefore
\begin{eqnarray*}
\langle u,xu\rangle &=&-\int^\infty_0dt\frac{d}{dt}e^{-\lambda
t}\langle W_tu,xW_tu\rangle\le-\int^\infty_0dte^{-\lambda
t}\frac{d}{dt}\|W_tu\|\\ &=&\|u\|^2
-\lambda\int^\infty_0dte^{-\lambda t}\|W_tu\|
=\lambda\int^\infty_0dte^{-\lambda t}\langle
u,(I-W^*_tW_t)\rangle\\ &=&\lambda\langle u,\tilde{\cal
E}^{(1)}_\lambda(I)u\rangle ,
\end{eqnarray*}
consequently $$0< x\le\lambda\tilde{\cal
E}^{(1)}_\lambda(I)=\lambda\ell_\lambda(I)+Q_\lambda(I).$$ If
$\ell_\lambda(I)\ne 0$ the proof is finished. In the case
$\ell_\lambda(I)=0$, from the above estimate we obtain $$0< x\le
Q_\lambda(I),$$ hence using (3.5) one gets for every $n\ge 1$,
$$0< x=Q^n_\lambda(x)\le Q^n_\lambda(I).$$ Therefore $$ 0<
x\le\lim_nQ^n_\lambda(I),$$ and this proves that
$\big(P^{\min}_t\big)_{t\ge 0}$ is non-conservative. \qed

\bigskip

The predual semigroup $(P^\dagger _t)_{t\ge 0}$ of a qds
$(P_t)_{t\ge 0}$ is the family of bounded operators on the Banach
space $({\cal T}(h), \|\cdot\|_1)$ of trace-class operators with
the trace norm $\|\rho\|_1=\tr |\rho|$, defined by means of the
relation $$\tr\big(P_t(x)\rho\big)=\tr\big(xP^\dagger
_t(\rho)\big),$$ for $x\in{\cal B}$ and $\rho\in{\cal T}(h)$.

If $\rho=|v\rangle\langle u|$ is the projector $(|v\rangle\langle
u|)\omega:=\langle u,\omega\rangle v$, for $u,v,\omega\in\h$, in
particular we have that $$\langle
u,P_t(x)v\rangle=\tr\big(P_t(x)|v\rangle\langle
u|\big)=\tr\big(xP^\dagger _t(|v\rangle\langle u|)\big),$$
$x\in{\cal B}$.

Since $(P_t)_{t\ge 0}$ is $\omega^*$-continuous, therefore
$P^\dagger_t$ is continuous with respect to the weak topology on
${\cal T}(h)$. Hence, by a well known result (see [1], Corollary
3.1.8, p. 168), $(P^\dagger _t)_{t\ge 0}$ is strongly continuous
and therefore a $C_0$-semigroup in ${\cal T}(h)$ and the weak and
strong generators coincide.

We need the following assumption on the $CP$ coefficient $\phi$ of
the formal generator ${\cal L}$.

\begin{itemize}
\item[(v)] There exists a Hilbert space $\ka$ with the inner product
$\langle\langle\cdot ,\cdot\rangle\rangle$, densely and
continuously included in $\h$, and $\phi(I)$ is a bounded
sesquilinear form on $\ka\times\ka$. Moreover we assume that
$D\subset\ka$.
\end{itemize}

From (v) and the Lax-Milgram Theorem it follows that there exists
a positive and self-adjoint operator $\Lambda$ on $\h$, with
$\dom\Lambda^{1/2}=\ka$ and $$\langle\langle
u,v\rangle\rangle=\langle\Lambda^{1/2}u,\Lambda^{1/2}v\rangle,$$
for any $u,v\in\h$. Furthermore we can assume $\Lambda\ge I$.

\medskip

On ${\cal T}(\h)$ let us consider the injective, contractive and
completely positive linear map $\beta :{\cal T}(\h)\to{\cal
T}(\h)$ defined by
$$\beta(\rho)=\Lambda^{-1/2}\rho\Lambda^{-1/2}.$$ A linear map
$\beta:{\cal T}(\h)\to{\cal T}(\h)$ is completely positive if
$$\sum_{i,j}\tr\big[x_ix_j\beta(\sigma_i\sigma^*_j)\big]\ge 0$$
for any pair of sequences $(x_j)\subset{\cal B}$,
$(\sigma_j)\subset{\cal T}_2(\h)$, where ${\cal T}_2$ is the space
of Hilbert-Schmidt operators in $\h$.

$\tilde{\cal T}=\beta({\cal T})$ will denote the range of $\beta$,
$\tilde{\cal T}$ has a natural structure of Banach space with the
norm $$\|\rho\|_{\tilde{\cal T}}:=\|\beta^{-1}(\rho)\|_1,
\quad\rho\in\tilde{\cal T}.$$ The map $\beta$ results to be an
isometric isomorphism from ${\cal T}$ onto $\tilde{\cal T}$.

We shall denote by ${\cal V}$ the subspace of ${\cal T}(h)$ of
rank-one operators $|v\rangle \langle u|, u,v\in D$. Since
$D\subset k$, it follows that ${\cal V}\subset\tilde{\cal T}$.

Assumption (v) was introduced in [3] in a different context. Some
important consequences of this assumption where studied in [3] and
[6]. In particular it was proved there that for any map $\phi$
satisfying (ii) and (v) there exists a map $\phi^\dagger
:\tilde{\cal T}\to{\cal T}$ contractive and completely positive
satisfying the relation $$\tr\big[x\phi^\dagger (|v\rangle\langle
u|)\big]=\phi(x)[u,v],$$ for any $x\in{\cal B}$ and $u,v\in D$.

An element $\rho\in{\cal T}(h)$ belongs to the domain $\dom{\cal
L}^\dagger $ of the generator ${\cal L}^\dagger $ of $(P^\dagger
_t)_{t\ge 0}$ if and only if there exists the limit $$\lim_{t\to
0^+}\frac{1}{t}\|P^\dagger _t(\rho)-\rho\|_1.$$

But, since the weak and strong generators coincide,
$\rho\in\dom{\cal L}^\dagger $ if and only if for every $x\in{\cal
B}$ the limit $$\lim_{t\to 0^+}\frac{1}{t}\tr\left(x(P^\dagger
_t(\rho)-\rho)\right)$$ exists.

The following is another criterion for explosion of a minimal qds.

\begin{prop}
If ${\cal L}$ is a formal generator satisfying (i)-(iii) and (iv)
with $\phi$ satisfying condition (v). Then the subspace ${\cal V}$
generated by the rank-one operators $|v\rangle\langle u|$, $u,v\in
D$ is contained in the domain $\dom{\cal L}^\dagger $ of the
generator of the predual semigroup $(P^{\min ,\dagger}_t)$ and
$${\cal L}\big(|v\rangle\langle u|\big)=\phi^\dagger
\big(|v\rangle\langle u|\big)-|v\rangle\langle
Gu|-|Gv\rangle\langle u|.$$

Moreover the following conditions are equivalent:
\begin{itemize}
\item[(i)] $(P^{\min}_t)_{t\ge 0}$ is non-conservative.
\item[(ii)] The orthogonal complement (or annihilator) in ${\cal
B}$ of $(\lambda-{\cal L}^\dagger )({\cal V})$ is non-trivial for
any $\lambda>0$.
\end{itemize}
\end{prop}

\medskip

\noindent{\bf Proof.} For $u,v\in D$ and every $x\in{\cal B}$ the
master equation (2.2) can be written in the form $$
\tr\big(P^{\min}_t(x)|v\rangle\langle u|\big) =\tr
\big(x|v\rangle\langle u|\big)+\int^t_0d\tau{\cal
L}(P^{\min}_t(x))[u,v].$$ Therefore we have
\begin{eqnarray*}
&&\frac{1}{t}\tr\left(x\left(P^{\min,\dagger}_t(|v\rangle\langle
u|)-|v\rangle\langle
u|\right)\right)=\frac{1}{t}\int^t_0d\tau\left(\phi(P^{\min}_\tau(x))[u,v]-\right.\\
&&\left.\langle Gu, P^{\min}_\tau(x)v\rangle -\langle u,
P^{\min}_\tau(x)Gv\rangle\right) =\\
&=&\frac{1}{t}\int^t_0d\tau\tr\left(P^{\min}_\tau(x)\left[\phi^\dagger
(|v\rangle\langle u|)-|v\rangle\langle Gu|-|Gv\rangle \langle
u|\right]\right)=\\ &=&\frac{1}{t}\int^t_0d\tau\tr\left(xP^{\min
,\dagger}_\tau\left(\phi^\dagger (|v\rangle\langle
u)-|v\rangle\langle-|Gv\rangle\langle u|\right)\right).
\end{eqnarray*}

From the weak continuity of $(P^{\min ,\dagger}_t)_{t\ge 0}$ we
obtain
\begin{eqnarray*}
&&\lim_{t\to 0^+}\frac{1}{t}\tr\left(x\left(P^{\min
,\dagger}_t(|v\rangle\langle u|)-|v\rangle\langle u|\right)\right)
=\\ &=&\tr\left(x\left(\phi^\dagger (|v\rangle\langle
u|)-|v\rangle\langle Gu|-|Gv\rangle\langle
u|\right)\right).\end{eqnarray*} This proves that ${\cal
V}\subset\dom{\cal L}^\dagger $ and $${\cal
L}^\dagger(|v\rangle\langle u|)=\phi^\dagger(|v\rangle\langle
u|)-|u\rangle\langle Gu|-|Gv\rangle\langle u|$$

To prove the equivalence of conditions (i) and (ii) observe that
$x$ is an element in the orthogonal complement in ${\cal B}$ of
$(\lambda -{\cal L}^\dagger )({\cal V})$ for some $\lambda>0$, if
and only if
\begin{eqnarray*}
0&=&\tr\left(x\left(\lambda-{\cal L}^\dagger
\right)\left(|v\rangle\langle
u|\right)\right)=\tr\left(x\left(\lambda|v\rangle\langle u|-{\cal
L}^\dagger \left(|v\rangle\langle u|\right)\right)\right)\\
&=&\tr\left(x\left(\lambda |v\rangle\langle u|-\phi^\dagger
\left(|v\rangle\langle u\right)+|v\rangle\langle
Gu|+|Gv\rangle\langle u|\right)\right) \\ &=& (\lambda -{\cal
L})(x)[u,v],
\end{eqnarray*}
for any $u,v\in D$. The result follows from the equivalence of
conditions (i) and (ii) in Theorem 3.3. \qed

\bigskip

The following Corollary is an extension of a criterion for
conservativity due to E. B. Davis (see [5], Prop. 3.3.2).

\begin{cor}
Assume that ${\cal L}$ is a formal generator satisfying (i)-(iii)
and (iv) with $\phi$ satisfying condition (v). Then the following
are equivalent
\begin{itemize}
\item[(i)] $(P^{\min}_t)_{t\ge 0}$ is conservative.
\item[(ii)] The subspace ${\cal V}$ of rank-one operators
$|v\rangle\langle u|$, $u,v\in D$ is a core for ${\cal
L}^\dagger$.
\end{itemize}
\end{cor}

\medskip

\noindent{\bf Proof.} The subspace ${\cal V}$ is dense in ${\cal
T}(\h)$. By Proposition 3.1 in [4], ${\cal V}$ is a core for
${\cal L}^\dagger$ if and only if $R(\lambda-{\cal
L}^\dagger)=(\lambda-{\cal L}^\dagger)({\cal V})$ is dense in
${\cal T}(\h)$ for some $\lambda >0$. This condition holds if and
only if the orthogonal complement (or anihilator) in ${\cal B}$ of
$(\lambda -{\cal L}^\dagger)({\cal V})$ is trivial for some
$\lambda>0$. The result follows from Proposition 3.4.

\section{Examples.}

\begin{exam}{\rm
On $\h=L_2(0,\infty)$ we shall consider operators induced by the
differential form $$\tau_fu=\frac{1}{2i}\left((fu)'+fu'\right),$$
where $f\in C^\infty(0,\infty)$, $f>0$, $f'$ is bounded and $\dps
\int^\infty_0dxf(x)^{-1}=\infty$. Notice that the function
$f(x)=(1+x)^\alpha$, $0\le\alpha\le 1$ satisfies these conditions.

We denote by $H_{1,0}$ the minimal operator induced by $\tau_f$,
it is defined by} $$\dom H_{1,0}=C^\infty_0(0,\infty)\quad{\rm
and}\quad H_{1,0}u=\tau_fu, \qquad u\in \dom H_{1,0}.$$
\end{exam}

The maximal operator $H_1$ induced by $\tau_f$ is defined by
$$\dom H_1=\{u\in L_2(0,\infty): u\;\;\mbox{is absolutely
continuous and}\;\;\tau_fu\in L_2(0,\infty)\}$$ and
$$H_1u=\tau_fu, \quad  u\in\dom H_1.$$

\bigskip

One can show that $H_0$ is a symmetric operator and that $H_{1,0}$
and $H_1$ are formal adjoints of each other. Moreover if $u\in\dom
H^*_{1,0}$ following [7], Theorem 6.29, pg. 160, one can see that
$u(x)=\omega(x)+cf^{1/2}(x),$ a.e. in $(0,\infty)$, where $\omega$
is absolutely continuous and $$\tau_f\omega=H^*_{1,0}u.$$
Therefore $u$ is absolutely continuous, and
$\tau_fu=\tau_f\omega=H^*_{1,0}u\in L_2(0,\infty)$; since $u\in
L_2(0,\infty)$, we can conclude that $u\in\dom H_1$ and this
proves that $H_{1,0}^*=H_1$.

Being symmetric the operator $H_{1,0}$ is closable and its closure
$\bar{H}_{1,0}$ is symmetric, moreover
$\bar{H}^*_{1,0}=H^*_{1,0}=H_1$. Notice that
$$\dom\bar{H}_{1,0}=\{u\in\dom H_1:\;\; u(0)=0\}.$$

Now let us consider the equations $$H^*_{1,0}u=\pm iu, \quad
u\in\dom H^*_{1,0}.$$ The solutions of these equations are
respectively
$$u_+(x)=c_1f(x)^{1/2}e^{-\int^x_0\frac{d\tau}{f(\tau)}}$$ and
$$u_-(x)=c_2f(x)^{-1/2}e^{+\int^x_0\frac{d\tau}{f(\tau)}}, \qquad
c_1,c_2\;\;{\rm nonzero}.$$ Notice that
$$\|u_+\|^2=c^2_1\int^\infty_0dxf(x)^{-1}e^{-2\int^x_0\frac{d\tau}{f(\tau)}}=\frac{1}{2}c^2_1<\infty,$$
since $\int^\infty_0\frac{d\tau}{f(\tau)}=\infty,$ therefore
$u_+\in L_2(0,\infty)$. Similarly one can show that $u_-\not\in
L_2(0,\infty)$. This proves that the defect indices of the
symmetric operator $\bar{H}_{1,0}$ are $n_+(\bar{H}_{1,0})=1$ and
$n_-(\bar{H}_{1,0})=0$.

\medskip

By the von Neumann Theorem we have that $$\dom
H_1=\dom\bar{H}_{1,0}\mathop{+}^\cdot N_+\mathop{+}^\cdot N_-$$
and $$H_1(\omega +v_++v_-)=\bar{H}_{1,0}\omega+iv_+-iv_-,$$
$\omega\in\dom\bar{H}_{1,0}$, $v_+\in N_+$, $v_-\in N_-$, where
$$N_+={\cal N}(i-H_1)={\cal R}(-i-\bar{H}_{1,0})^\perp\;\;{\rm
and}\;\; {\cal N}_-=N(-i-H_1)=i={\cal R}(i-\bar{H}_{1,0})^\perp$$
are the defect subspaces of $\bar{H}_{1,0}$ and
$\dps\mathop{+}^{\cdot}$ denotes direct sum.

\medskip

But we have shown that $N_-=\{0\}$, therefore $$\dom
H_1=\dom\bar{H}_{1,0}\mathop{+}^{\cdot} N_+$$ and $$H_1(\omega
+v_+)=\bar{H}_{1,0}\omega +iv_+,$$ $\omega\in\dom\bar{H}_{1,0}$,
$v_+\in N_+$.

Then for $u\in\dom H_1$, $u=\omega +v_+$,
$\omega\in\dom\bar{H}_{1,0}$, $v_+\in N_+$ we have that $$\langle
iH_1u,u\rangle =-i\langle\bar{H}_{1,0}\omega, \omega\rangle
+2i{\rm Im}\langle\omega ,v_+\rangle -\|v_+\|^2,$$ hence
$$Re\langle iH_1u, u\rangle =-\|v_+\|^2\le 0.$$ This proves that
$iH_1$ is dissipative.

If $\Theta(iH_1)=\{\langle iH_1u,u\rangle : u\in\dom
H_1,\;\|u\|=1\}$ is the numerical range of $iH_1$, then we have 0
for $\lambda_0>0$ $$\delta =\;{\rm
dist.}\big(\lambda_0,\overline{\Theta(iH_1)}\big)>0.$$ Therefore
\begin{eqnarray*}
\delta&\le&\big|\langle iH_1u, u\rangle
-\lambda_0\big|=\big|\langle (iH_1-\lambda_0I)u,u\rangle\big|\le\\
&\le& \|(iH_1-\lambda_0I)u\|,
\end{eqnarray*}
for any $u\in\dom H_1$, $\|u\|=1$. Hence the operator
$(iH_1-\lambda_0I)^{-1}$ there exists, it is bounded and closed on
${\cal R}(iH_1-\lambda_0I)$. Then $R(iH_1-\lambda_0)$ is closed
and therefore $${\cal R}(iH_1-\lambda_0I)=\h ,$$ since ${\cal
R}(iH_1-\lambda_0(I)^\perp =\{0\}$ for any $\lambda_0>0$.

By the Lumer-Phillips Theorem we conclude that $-G=iH_1$ is the
generator of a $C_0$-semigroup of contractions $(W_t)_{t\ge 0}$ in
$\h$.

\medskip

Let us consider the formal generator ${\cal L}$ that associates
with every element $x\in{\cal B}={\cal B}\big(L_2(0,\infty)\big)$
the sesquilinear form $${\cal L}(x)[u,v]=-\langle Gu,xv\rangle
-\langle u,xGv\rangle,\eqno(4.1)$$ $u,v\in\dom G$. In this case
the CP part of ${\cal L}$ is zero, $\phi(x)=0$, $x\in{\cal B}$.

\bigskip

${\cal L}$ satisfies conditions (i)-(iii) in section 3 and for
$u,v\in D=\dom\bar{H}_{1,0}\subsetneq\dom H_1=\dom G$ we have that
\begin{eqnarray*}
{\cal L}(I)[u,v]&=&\langle i\bar{H}_{1,0}u,v\rangle+\langle u,i
\bar{H}_{1,0}v\rangle =\\ &=& i\big(-\langle
\bar{H}_{1,0}u,v\rangle +\langle
u,\bar{H}_{1,0}v\rangle\big)=0,\end{eqnarray*} since $H_{1,0}$ is
symmetric. Hence ${\cal L}$ satisfies also condition (iv) in the
previous section with $D=\;\dom\bar{H}_{1,0}$.

\medskip
Since $\phi=0$ we have that $\dps\lim_nQ^n_\lambda(I)=0$. If
$\ell_\lambda(I)=0$, we have from (4.1) that for every $u\in\dom
G$, $\dps\int^t_0d\tau{\cal L}(I)(W_{t-\tau})=0$ a.e. $t\ge 0$ and
taking derivative we obtain ${\cal L}(I)[u]=0$ for all $u\in\dom
G$. This implies that $H_1$ is symmetric, but we know that
$\bar{H}_{1,0}$ is maximal symmetric and
$\bar{H}_{1,0}\varsubsetneq H_1$. Therefore $\ell_\lambda(I)\ne 0$
and the minimal semigroup constructed from the formal generator
${\cal L}$ is non-conservative.

\medskip
The minimal qds constructed from the formal generator (4.1) is
$$P^{\min}_t(x)=W^*_txW_t,\quad x\in{\cal B}.$$ Observe that
$(P^{\min}_t)_{t\ge 0}$ is conservative $\big(P^{\min}_t(I)=I,
\;t\ge 0\big)$, if and only if the $C_0$-semigroup $(W_t)_{t\ge
0}$ is a semigroup of isometries: $\|W_tu\|=\|u\|$, $u\in\h$.

\begin{exam} {\rm The adjoint semigroup $(U_t)_{t\ge 0}$ defined
by $U_t\!=\!W^*_t$, $t\!\ge\! 0$ is a strongly continuous
semigroup of contractions with the infinitesimal generator
$(-G)^*=-iH^*_1=-i\bar{H}_{1,0}$. The associated minimal qds
$$P^{\min}_t(x)=U^*_txU_t, \;\; t\ge 0, \; x\in{\cal B},$$ is
conservative, because its formal generator is defined by $${\cal
L}(x)[u,v]=-\langle i\bar{H}_{1,0}u, xv\rangle -\langle
u,xi\bar{H}_{1,0}v\rangle .$$ for $x\in{\cal B}$,
$u,v\in\dom\bar{H}_{1,0}$. Therefore we have $\phi(x)=0$,
$x\in{\cal B}$ and hence $\dps\lim_nQ^n_\lambda(I)=0$; moreover
for $u\in\dom\bar{H}_{1,0}$
$$\ell_\lambda(I)[u]=\int^\infty_0dte^{-\lambda
t}\int^t_0d\tau{\cal L}(I)[W_{t-\tau}u]=0,$$ since $\bar{H}_{1,0}$
is symmetric.}
\end{exam}

\begin{exam} {\rm Take $\h$ and $H_1$ as in Example 4.1 and consider the
CP map that associates with every element $x\in{\cal B}$ the
sesquilinear form defined for $u,v\in\dom H_1$ by
$$\phi(x)[u,v]=\langle Lu, xLv\rangle,$$ where $L$ is the operator
of multiplication by a complex-valued function $\ell(s)$,
$s\in(0,\infty)$. Then we have that $$\phi(I)[u,v]=\langle u,
|\ell|^2v\rangle,$$ $u,v\in\dom H_1$, i.e., $\phi(I)$ coincides
with the operator of multiplication by the positive function
$|\ell(s)|^2$, $s\in(0,\infty)$.}
\end{exam}

Assume that $-G=-\frac{1}{2}\phi(I)+iH_1$, with $\dom G=\dom H_1$,
is the generator of a strongly continuous semigroup of
contractions in $h$, $(W_t)_{t\ge 0}$, and let us consider the
formal generator ${\cal L}$ that associates with every element
$x\in{\cal B}$ the sesquilinear form $${\cal
L}(x)[u,v]=\phi(x)[u,v]-\langle Gu, xu\rangle-\langle u,xGv\rangle
,$$ $u,v\in\dom G$.

${\cal L}$ satisfies conditions (i) and (ii) in Section 2,
moreover
\begin{eqnarray*}
{\cal
L}(I)[u]&=&\phi(I)[u]-\left\langle\left(\frac{1}{2}\phi(I)-iH_1\right)u,u\right\rangle
\\ &&-\left\langle
u,\left(\frac{1}{2}\phi(I)-iH_1\right)u\right\rangle =\langle
u,|\ell|^2u\rangle-\frac{1}{2}\langle|\ell|^2u,u\rangle \\
&&+\langle iH_1u,u\rangle -\frac{1}{2}\langle
u,|\ell|^2u\rangle+\langle u,iH_1u\rangle \\ &=& 2Re\langle
iH_1u,u\rangle\le 0,\end{eqnarray*} since $iH_1$ is dissipative.
Hence ${\cal L}$ satisfies also condition (iii) in Section 2.

Notice that for $u\in\dom\bar{H}_{1,0}$ we have that $${\cal
L}(I)[u]=2Re\langle iH_1u,u\rangle =2Re\langle
i\bar{H}_{1,0}u,u\rangle =0,$$ since $\bar{H}_{1,0}$ is symmetric.
Therefore ${\cal L}$ satisfies our condition (iv) in the previous
section with $D=\dom\bar{H}_{1,0}$.

The minimal qds constructed from this formal generator ${\cal L}$
is non-conservative because as in Example 4.1, $\ell_\lambda(I)=0$
implies that $H_1$ is symmetric but $\bar{H}_{1,0}\varsubsetneq
H_1$ and $\bar{H}_{1,0}$ is maximal symmetric.

\bigskip

To observe the connection of the conservativity problem for formal
generators (4.1) with the von Neumann theory of the defect indices
of a symmetric operator, we prove the following.

\medskip

\begin{prop} Let ${\cal L}$ be the formal generator given by
equation (4.1). Then the following conditions are equivalent
\begin{itemize}
\item[(i)] The defect index $n_+(\bar{H}_{1,0})$ of the closed symmetric
operator $$\bar{H}_{1,0}=iG|_D\;\mbox{is positive},\;
n_+(\bar{H}_{1,0})>0.$$
\item[(ii)] The equation
$${\cal L}(x)[u,v]=\lambda\langle u,xv\rangle, \quad
u,v\in\dom\bar{H}_{1,0},$$ has a positive, bounded solution
$x\in{\cal B}$ for some $\lambda>0$.
\end{itemize}
\end{prop}

\noindent{\bf Proof.} Assume that $N_+(\bar{H}_{1,0})={\cal
N}(i-H^*_{1,0})\ne\{0\}$ and take $u\in N_+(\bar{H}_{1,0})$, $u\ne
0$. Let $x\in{\cal B}$ be the projector $x=|u\rangle\langle u|$,
then we have for every $v\in\dom\bar{H}_{1,0}$ that
\begin{eqnarray*}
&&{\cal L}(x)[v]=\langle i\bar{H}_{1,0}v,|u\rangle\langle
u|v\rangle+\langle v,|u\rangle\langle u|i\bar{H}_{1,0}v\rangle \\
&=& -i\langle v,H^*_{1,0}u\rangle\langle u,v\rangle+i\langle
v,u\rangle\langle H^*_{1,0}u,v\rangle\\ &=& -i\langle u,
iu\rangle\langle u,v\rangle +\langle v,u\rangle\langle iu,v\rangle
=2\langle u,v\rangle\langle v,u\rangle \\ &=& 2\langle v,xv\rangle
.\end{eqnarray*}

Using the polarization identity we obtain that $${\cal
L}(x)[u,v]=2\langle u,xv\rangle, \qquad u,v\in\dom\bar{H}_{1,0}.$$
Therefore (ii) holds with $\lambda=2$  if $n_+(H_{1,0})>0$.

Conversely, assume that (ii) holds and $n_+(\bar{H}_{1,0})=0$,
i.e., $N_+(\bar{H}_{1,0})={\cal N}(i-H^*_{1,0})=\{0\}$. Therefore
$$ {\cal R}(I-i\bar{H}_{1,0})^\perp={\cal
R}(-i-\bar{H}_{1,0}^*)^\perp ={\cal N}(i-\bar{H}^*_{1,0})=\{0\}.$$
Take $u\in\dom G={\cal R}\big((I+G^{-1})\big)$ and let $v=(I+G)u$.
Since $(I-i\bar{H}_{1,0})\dom\bar{H}_{1,0}={\cal R}(I-iH_{1,0})$
is dense in $\h$, for any $\epsilon
>0$ there exists $v_\epsilon=(I-i\bar{H}_{1,0})u_\epsilon
=(I+G)u_\epsilon$, $u_\epsilon\in\dom\bar{H}_{1,0}$, such that
$\|v-v_\epsilon\|<\epsilon$. Therefore we have that
$$\|u-u_\epsilon\|^2=\big\|(I+G)^{-1}v-(I+G)^{-1}v_\epsilon\big\|\le
\|v-v_\epsilon\|<\epsilon,$$ by the Hille-Yosida Theorem.

Then we have proved that $u_\epsilon\in\dom{\bar H}_{1,0}$,
$u_\epsilon\to u$ and $(I-i\bar{H}_{1,0})u_\epsilon\to (I+G)u$,
hence $-i\bar{H}_{1,0}u_\epsilon\to Gu$, as  $\epsilon\to 0$. This
implies that $u\in\dom\bar{H}_{1,0}$ and hence
$iH_1=-G=i\bar{H}_{1,0}$, i.e., $H_1=\bar{H}_{1,0}$.

The relation ${\cal L}(x)[u,v]=\lambda\langle u,xv\rangle$ holds
for $u,v\in\dom{\bar H}_{1,0}$ and some $\lambda >0$, therefore
for every $t\ge 0$ and $u,v\in \dom G$, we have that
$$-\lambda\langle W_tu,xW_tv\rangle-\langle
GW_tu,xW_tv\rangle-\langle W_tu,xGW_tv\rangle =0.$$ Equivalently
we have that $$\frac{d}{dt}e^{-\lambda t}\langle W_tu,xW_tv\rangle
=0,$$ and integrating we obtain
$$0=\int^\infty_0\frac{d}{dt}e^{-\lambda t}\langle
W_tu,xW_tv\rangle = \langle u,xv\rangle,$$ for all $u,v\in\dom G$.
Then $x=0$ and this finishes the proof.\qed

\bigskip

Since $n_+(\bar{H}_{1,0})>0$, Proposition 4.2 and Theorem 3.3 give
another proof that the minimal qds of Example 4.1 is
non-conservative or explosive.

\bigskip

The above proposition holds in the case when $-G=i H^*$, with $H$
any maximal symmetric closed operator in a Hilbert space $\h$. It
says that in the case when $iH$ is the restriction of a generator
of a strongly continuous semigroup of contractions $(W_t)_{t\ge
0}$ in $\h$, then $(W_t)_{t\ge 0}$ is a semigroup of isometries if
and only if $n_+ (H)=0$.

Given a closed symmetric operator $H$ it naturally arises the
question of whether or not is $iH$ the restriction of a generator
of a strongly continuous semigroup isometries. The following
proposition give an answer.

\begin{prop} Let $H$ be a closed symmetric operator in a Hilbert
space $\h$ with finite defect indices $(n_+, n_-)$, then
\begin{itemize}
\item[(i)] if $n_+\le n_-$ the operator $iH$ is the restriction of
a generator of a strongly continuous semigroup isometries in $\h$.
\item[(ii)] if $n_+>n_-$ then $iH$ is not the restriction of a
generator of a strongly continuous semigroup of isometries in
$\h$.
\end{itemize}
\end{prop}

\medskip

\noindent{\bf Proof.} (i) If $0=n_+\le n_-$ then $H$ is maximal
symmetric (or selfadjoint if $n_-=0$). Therefore the arguments in
Example 4.1 help to prove that $-iH^*$ generates a strongly
continuous semigroup of contractions in $\h$. The adjoint
semigroup $(U_t=W^*_t)_{t\ge 0}$ is generated by $iH$, since
$n_+=0$, this semigroup $(U_t)_{t\ge 0}$ is of isometries by
Proposition 4.4. If $H$ is selfadjoint, $iH$ generates a unitary
group.

If $0<n_+\le n_-$, then the defect subspace $N_+$ of $H$ is
isometrically isomorphic with a subspace $F_-$ of $N_-$, let us
denote by $V$ the isometry $V:N_+\to F_-$. By the von Neumann
Theorem, associated with $V$ there exists a closed symmetric
extension $H_V$ of $H$ defined as $$\dom H_V=\dom H +\{v+Vv:\;
v\in N_+\}$$ and $$H_V(u+v+Vv)=H_u+iv-iVv=H^*(u+v+Vv)$$ for
$u\in\dom H$ and $v\in N_+$.

Since ${\cal R}(-i-H_V)={\cal R}(-i-H)\mathop{+}^{\cdot} N_+={\cal
R}(-i-H)\mathop{+}^{\cdot}{\cal R}(-i-H)^\perp=\h$ we have that
$n_+(H_V)=0$, hence we are in the case $0=n_+(H_V)\le n_-(H_V)$.
So we can proceed as above to prove that $iH_V$ generates a
$C_0$-semigroup of isometries in $\h$, and hence $iH$ is the
restriction of a generator of a strongly continuous semigroup of
isometries in $\h$.

\medskip

\noindent(ii) If $n_+>n_-$ then there exists an isometry $V'$ from
the defect subspace $N_-$ of $H$ onto a proper subspace of $N_+$,
and associated with $V'$ exists a maximal symmetric extension
$H_{V'}$ of $H$. The semigroup of contractions generated by the
disipative operator $iH^*_{V'}$ is not a semigroup of isometries
since $n_+(H_{V'})>0$.\qed

\medskip

\begin{exam} {\rm In $h=\ell_2(\k)$, with the complete orthonormal
system $(e_n)_{n\ge 0}$, let $V$ be the isometry defined by
$$Ve_n=e_{n+1}. \qquad n\ge 0.$$ So we have that $D(V)=\h$ and}
${\cal R}(V)=\span\{e_n,\;n\ge 1\}$.
\end{exam}

Therefore from the von Neumann Theorem, there exists a symmetric
operator $H$ given by the Cayley transform $$H=i(I+V)(I-V)^{-1},$$
if and only if ${\cal R}(I-V)$ is dense in $\h$. But $v\in {\cal
R}(I-V)^\perp$ implies that $$\langle v,e_n-e_{n+1}\rangle =0,
\qquad n\ge 0,$$ hence $$0=\sum^{n-1}_{k=0}\langle
v,e_k-e_{k+1}\rangle=\langle v,e_0\rangle -\langle v,e_n\rangle,
\quad n\ge 1,$$ or $$\langle v,e_0\rangle =\langle
v,e_n\rangle,\qquad n\ge 1.$$ This implies that $v=0$ and hence
${\cal R}(I-V)$ is dense in $\h$.

\medskip

The isometry $V$ is closed, therefore $H$ is closed and $\dom
V={\cal R}(i+H)$, ${\cal R}(V)={\cal R}(I-H)$. Hence we obtain
$$N_+(H)={\cal R}(-i-H)^\perp=\{0\}\quad{\rm and}\quad
N_-(H)={\cal R}(i-H)^\perp =\span\{e_0\}.$$ Then $n_+(H)=0$ and
$n_-(H)=1$. By Proposition 4.4 $iH$ is not the restriction of a
generator of a $C_0$-semigroup of isometries.

A similar result is obtained when $V_m$ is the isometry defined by
$$V_me_n=e_{n+m}, \quad n\ge 0\quad{\rm and}\quad m>1 \;\;{\rm
fixed}.$$

\end{document}